\documentclass{article}
\usepackage{spconf,amsmath,graphicx}

\usepackage{booktabs, multirow} 
\usepackage{soul}
\usepackage[table]{xcolor} 
\usepackage{changepage,threeparttable} 
\usepackage{multirow}
\usepackage{amssymb}
\usepackage{float}

\title{SpatialEmb: Extract and Encode Spatial Information for 1-Stage Multi-channel Multi-speaker ASR on Arbitrary Microphone Arrays}
\name{Yiwen Shao$^{1,2}$\sthanks{This work was done while Yiwen was a research intern at Tencent AI Lab, USA}, Yong Xu$^2$, Sanjeev Khudanpur$^1$, Dong Yu$^2$}
\address{$^1$Center for Language and Speech Processing, Johns Hopkins University, Baltimore, MD, USA\\
  $^2$Tencent AI Lab, Bellevue, WA, USA}
\begin{document}
%
\maketitle
\begin{abstract}
Spatial information is a critical clue for multi-channel multi-speaker target speech recognition. Most state-of-the-art multi-channel Automatic
Speech Recognition (ASR) systems extract spatial features only during the speech separation stage, followed by standard single-channel ASR on the separated speech. This approach results in an inefficient, lengthy pipeline and sub-optimal ASR performance due to the accumulated errors from preprocessing modules. Furthermore, most spatial feature extraction methods depend on the knowledge of speaker positions and microphone topology, making the systems reliant on specific settings and challenging to adapt to new equipment. In this work, we propose a solution to these issues with a lightweight embedding module named \textit{SpatialEmb}, which extracts and encodes spatial information directly for the ASR model, supporting both fixed and arbitrary microphone topology. We conduct comprehensive experiments on AliMeeting, a real meeting corpus, to determine the optimal model design for \textit{SpatialEmb} in terms of both performance and efficiency. Our best model trained with 105 hours \textit{Train-Ali-far} achieves 17.04\% and 20.32\% character error rates (CER) on the Eval and Test sets, establishing a new state-of-the-art result with the same training data.

\end{abstract}
\section{Introduction}
\label{sec:intro}

\begin{figure}
    \centering
    \includegraphics[width=80mm]{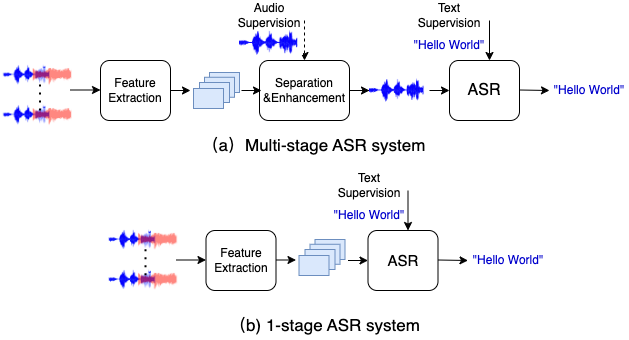}
    \caption{An illustration of (a) Multi-stage ASR system and (b) 1-stage ASR system}
    \label{fig:system}
\end{figure}

Multi-channel multi-speaker automatic speech recognition (ASR) is particularly challenging due to overlapping speech. Existing approaches are categorized into multi-stage and 1-stage systems (Fig. \ref{fig:system}).

Multi-stage systems first apply speech preprocessing, like guided source separation (GSS) \cite{kanda2019guided, raj2022gpu}, before passing the output to an ASR module. This preprocessing can be offline or end-to-end with the ASR module \cite{9984940, li2021mimo, yu2021audio, zhang2021adl, masuyama2023end}, but often requires additional pre-training and is computationally intensive, limiting real-time application suitability \cite{raj2022gpu}.

1-stage systems bypass explicit preprocessing, feeding multi-channel overlapped speech directly into the ASR module \cite{shao2022multi, yu2023mfcca}. These systems simultaneously generate multiple transcripts using techniques like permutation invariant training (PIT) \cite{yu2017permutation} or leverage spatial features from microphone arrays to enhance recognition without extra speaker registration \cite{chen2018multi}. Recent advancements \cite{shao2022multi, shao2023rir} have achieved state-of-the-art performance by utilizing spatial features derived from room impulse responses (RIR), without requiring visual input or specific microphone knowledge \cite{shao2024multi}. While some studies explore multi-channel ASR with arbitrary microphone arrays, they often incur high computational costs \cite{luo2020end, yoshioka2022vararray, huang2024unix}.

This work proposes a streamlined 1-stage multi-channel ASR module with minimal modifications to existing single-channel models. Using Conformer \cite{gulati2020conformer} as the ASR backbone, we concatenate spectral features with Solo-SF and evaluate three embedding structures: \textit{Conv2d}, \textit{ConvNext} \cite{liu2022convnet, jiang2022nextformer}, and \textit{GRU-Conv2d}. Experiments on the AliMeeting corpus \cite{yu2021audio} identify the optimal structures for performance and efficiency. We also introduce a parameter-free \textit{divide-average-concate} (DAC) method to support arbitrary microphone arrays, outperforming existing methods with lower computational costs.

Integrating these enhancements into \textit{SpatialEmb}, our best model achieves 17.04\% and 20.32\% CER on the \textit{Eval} and \textit{Test} sets, respectively, setting a new state-of-the-art on the AliMeeting corpus with 105 hours of \textit{Train-Ali-far} data.

\section{Feature Extraction: Spectral and Spatial Features}
\subsection{Spectral feature}
We begin with the short-time Fourier transform (STFT) complex coefficient \( Y \in \mathbb{C}^{M \times T \times F} \) of the overlapped speech, where \( M \), \( T \), and \( F \) represent the channel, time, and frequency dimensions, respectively. The spectral feature is derived from the magnitude of \( Y \). One common choice is to use the log power spectrum (LPS) \( \in \mathbb{R}^{M \times T \times F} \):

\begin{equation}
    \text{LPS} = \log(|Y|^2)
    \label{lps}
\end{equation}

Another widely used spectral feature for ASR is the log filter bank (LFB) \( \in \mathbb{R}^{M \times T \times N} \). It is obtained by applying a bank of predefined filters \( \text{FB} \in \mathbb{R}^{F \times F'} \) to the power spectrum of \( Y \), where \( F' \) is the number of filter bank bins:

\begin{equation}
    \text{LFB} = \log(|Y|^2 \times \text{FB})
    \label{fbank}
\end{equation}

In this work, we use a 201-dimensional LPS (as described in Equation \ref{lps}) with a 25 ms window length and a 10 ms shift size as our baseline spectral feature. We will also compare it with the 80-dimensional LFB later in the experiments section.

\subsection{Spatial feature}
To distinguish the target speech in the overlapped part, the phase difference within different channels of $Y$ can be utilized. \cite{shao2023rir} exploit the underlying room impluse response (RIR) of the target speaker to the microphone array, and propose to convolve the overlapped speech $Y$ with a short kernel of either an explict RIR or a solo segment implicitly containing the same RIR. This process defines an intermediate phase known as the RIR-convolved phase (RP) $\in \mathbb{R}^{M\times T \times F}$:
\begin{align}
    \text{RP}_{m,t,f} &= \angle \left( Y * R^H \right)_{m,t,f} \\
    &= \angle \left( \sum_{k=0}^{K-1} Y_{m,t-k, f} \cdot R^H_{m,k,f} \right)
\end{align}
where $\angle()$ is the operation of taking the phase and $()^H$ denotes the conjugate of a complex number. $R \in \mathbb{C}^{M \times T \times K}$ is the $K$-frame kernel from either the first $K$ frames of the target RIR or a $K$-frame segment from the solo part of the target speaker. As the solo segment is more robust and accessible, we are using \textbf{solo segment} as the kernel for this work. We take $K = 10$ which equals to 0.1 second of solo segment. The best setting of \cite{shao2023rir,shao2024multi} is used, where the solo segment is selected from a 2-second \textbf{nearest} solo part using \textbf{composition} method. The solo part can either be obtained from utterance level diarization, or just from preceding solo part in a long recording. We assume such information is accessible in this work without distraction from the main focus.

If $Y$ and $R$ share the same underlying RIR at specific T-F bins, which means $Y$ is dominated by the target speaker, similar pattern in RP will be shown across channels. Then we can extract the pair-wise spatial feature (SF) $\in \mathbb{R}^{T \times F}$ as the cosine of interchannel RP differences:
\begin{equation}
    \text{SF}_{t,f}(m_1, m_2) = \cos (\text{RP}_{m_1, t,f} - \text{RP}_{m_2, t,f}) \\
    \label{sf_pair}
\end{equation}
To make SF independent to pre-defined $(m_1,m_2)$ microphone pairs, we modify Equation \ref{sf_pair} to consider all pairs of microphones:
\begin{equation}
    \text{SF}_{t,f} = \frac{1}{M(M-1)}\sum_{i=1:M} \sum_{j \neq i} \cos ( \text{RP}_{i, t,f} - \text{RP}_{j, t,f})
\end{equation}
In this way, $\text{SF}_{t,f}$ serve similarly to a T-F mask, which indicates the dominance of the target speaker in the overlapped speech $Y$.

\subsection{Feature fusion}
\subsubsection{Fixed microphone topology setting}
\label{fixed}
For fixed topology setting, we simply concatenate the multi-channel spectral feature of shape $[M, T, F]$ and spatial feature of shape $[1, T, F]$ on the channel dimension, resulting in a final input of shape $[M+1, T, F]$. Specially, if the 80-dim LFB is used as spectral feature, to make the spatial feature has the same dimension for concatenation, we apply the same filter bank as in Equation \ref{fbank} on the spatial feature as well.
\subsubsection{Arbitrary microphone topology setting}
For arbitrary microphone topology, we have 2 settings for different approaches that we will compare in the following sections:
\begin{enumerate}
    \item \textbf{Spectral Feature Squeezing:} we fuse the multi-channel spectral feature to a single channel feature of shape $[T, F]$ and then concatenate it with the spatial feature, resulting in a fixed input shape of $[2, T, F]$;
    \item \textbf {Spatial Feature Expansion:} we make $M$ copies of spatial feature, and then concatenate each of them to the corresponding spectral feature, resulting in a 4D input tensor of shape $[M, 2, T, F]$;
\end{enumerate}

\section{Systems for comparison}
\subsection{Multi-stage system}
\subsubsection{Pipeline system}
\textbf{Guided Source Separation (GSS) \cite{kanda2019guided}} is a widely used offline speech separation approach for multi-stage, multi-channel, multi-speaker ASR systems. Initially proposed during the CHiME-5 challenge \cite{barker2018fifth}, GSS uses pre-computed speaker activities (i.e., diarization) to estimate time-frequency masks for the target speaker using a complex angular central Gaussian mixture model (CACGMM) \cite{ito2016complex}. These masks are used in mask-based MVDR beamforming to isolate single-channel target speech from the mixture. The ASR model is then trained and tested on the GSS-enhanced speech.

Despite achieving remarkable results, GSS is limited by significant computational cost. Although a GPU-accelerated version exists \cite{raj2022gpu}, it still requires offline preprocessing and relies heavily on diarization performance, making it less suitable for real-time applications. In this work, we provide GSS with ground truth diarization to establish it as a strong competing system.

\subsubsection{End-to-end system}
\textbf{All deep learning MVDR beamformer (ADL-BF) \cite{zhang2021adl}} and its variants \cite{li2021mimo, xu2021generalized} are widely adopted in end-to-end multi-speaker ASR systems. These systems can be divided into a separation stage and an ASR stage. The separation stage consists of a complex-valued ratio filter (cRF) estimator from spectral and spatial features and a neural beamformer that learns the beamforming weights from the spatial covariance matrices (SCMs):

\begin{align}
    W &= \text{NN}([\Phi_{ss}, \Phi_{nn}]) \\
    X_{t,f} &= W_{t,f}^H \cdot Y_{t,f}
\end{align}

where \(\Phi_{ss}, \Phi_{nn} \in \mathbb{C}^{M^2 \times T \times F}\) are the flattened SCMs for the target speech and noise, respectively. \(W_{t,f} \in \mathbb{C}^M\) is the estimated beamforming vector, and \(X_{t,f} \in \mathbb{C}\) is the predicted STFT coefficient for the separated target speech. We use the \textit{Linear-GRU-GRU-Linear} model structure for NN \cite{li2021mimo}.

To further improve the ADL-BF baseline, we apply the narrow-band approach introduced in \cite{quan2022multi}, estimating frequency-wise beamforming weights \(W_f \in \mathbb{C}^{M \times T}\) using a shared network:

\begin{equation}
    W_f = \text{NN}([\Phi_{ss, f}, \Phi_{nn, f}])
\end{equation}

The neural network architecture remains the same, with the input being the SCMs for each frequency. This process can be done in parallel, treating the frequency dimension as the batch dimension. Consequently, the computational cost increases by a factor of \(F\).

Since ADL-BF also uses spatial features as input, we apply the same input features as described in Section \ref{fixed} for a fair comparison. The output of ADL-BF is single-channel separated speech \(X\) and can be trained end-to-end with ASR loss. Thus, this system can also be viewed as a 1-stage system if there is no need for intermediate separation supervision.

\subsection{1-stage system}
The attention mechanism is widely used for 1-stage multi-speaker ASR systems \cite{chang2021end, chang2021multi}. Among them, the recently developed \textbf{Multi-frame Cross-Channel Attention (MFCCA) \cite{yu2023mfcca}} achieves state-of-the-art results on the AliMeeting dataset. This method integrates MFCCA into repeated Conformer blocks without requiring extra preprocessing modules. All intermediate representations of the speech are of shape \([M, T, D]\), where \(D\) is the embedding dimension, eliminating the need to merge multi-channel information early.

At the end, convolutional fusion blocks integrate the multi-channel tensors into a standard one of shape \([T, D]\). The remainder of the system, including the decoder, functions similarly to a regular single-channel ASR system. However, this method must maintain a 3D representation for most of the model, leading to significantly higher computational costs.

\section{Proposed: SpatialEmb}
\subsection{System Overview}
\begin{figure}
    \centering
    \includegraphics[width=80mm]{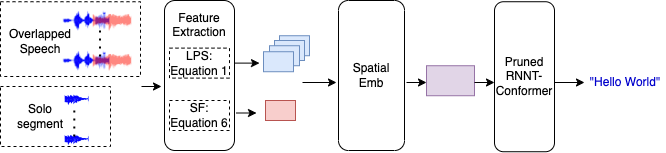}
    \caption{System overview of the proposed 1-stage ASR system using \textit{SpatialEmb}.}
    \label{fig:overview}
\end{figure}

We propose handling multi-channel information in the feature extraction and embedding modules within a 1-stage ASR system, leaving all other parts unchanged. As shown in Fig. \ref{fig:overview}, the system takes multi-channel overlapped speech and a target solo segment as input. The feature extraction module outputs multi-channel spectral and spatial features, following Equations 1 and 6. These features are fused as described in Section 2.3 and sent to the \textit{SpatialEmb} module to extract the final embeddings. The rest of the system is a standard Conformer with pruned RNNT loss \cite{kuang2022pruned}, implemented in Icefall\footnote{\label{icefall}https://github.com/k2-fsa/icefall}.

\subsection{SpatialEmb Module Structure}
\begin{figure}
    \centering
    \includegraphics[width=80mm]{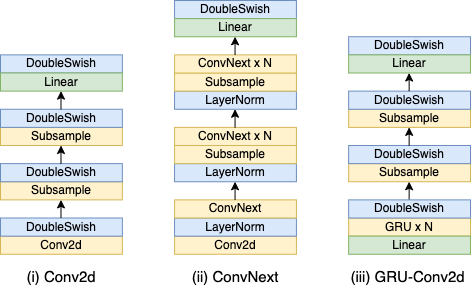}
    \caption{Three types of \textit{SpatialEmb} module structures.}
    \label{fig:emb}
\end{figure}

We examine three types of \textit{SpatialEmb} module structures (Fig. \ref{fig:emb}):

\begin{enumerate}
    \item \textit{Conv2d}: A Conv2d layer with a kernel size of \(3 \times 1\) (T $\times$ F) and a stride of 1.
    \item \textit{Subsample}: A Conv2d layer with a kernel size of \(3 \times 3\) and a stride of \(2 \times 2\).
    \item \textit{ConvNext} \cite{liu2022convnet}: A ConvNext block\footnote{https://github.com/facebookresearch/ConvNeXt} consisting of a depthwise Conv2d layer (\(7 \times 7\)), a LayerNorm layer, a pointwise Conv2d layer (\(1 \times 1\)), a GeLU activation layer, and another pointwise Conv2d layer (\(1 \times 1\)).
    \item \textit{DoubleSwish}: An activation layer that approximates \(\text{Swish}(\text{Swish}(x))\), implemented in Icefall.
\end{enumerate}

\subsubsection{Conv2d}
\vspace{-0.2cm}
The \textit{Conv2d} (Fig. \ref{fig:emb} i) structure is the standard embedding layer for a Conformer network and serves as the baseline. We use the Icefall implementation, increasing the number of input channels in the first \textit{Conv2d} layer from 1 to \(M+1\).

\subsubsection{ConvNext}
\vspace{-0.2cm}
The \textit{ConvNext} \cite{liu2022convnet} (Fig. \ref{fig:emb} ii) is a state-of-the-art convolutional model for computer vision, also applied to many ASR models \cite{jiang2022nextformer, yao2023zipformer}. We use the standard \textit{ConvNext} block and maintain an overall downsampling factor of 4 using two \textit{Subsample} layers. The number of \textit{ConvNext} blocks, denoted as \(N\), will vary in the experiments section.

\subsubsection{GRU-Conv2d}
\vspace{-0.2cm}
Inspired by the \( \textit{Linear-GRU} \) structure in ADL-BF (Section 3.1.2), we replace the first \textit{Conv2d} layer in the baseline structure with a \( \textit{Linear} \) layer followed by \(N\) stacks of \( \textit{GRU} \) layers (Fig. \ref{fig:emb} iii). This maps the input feature \([M+1, T, F]\) to an intermediate representation \([C, T, F]\), where \(C\) is the input channel number of the subsequent \textit{Subsample} layer. All other parts remain unchanged.

\subsection{Arbitrary Microphone Topology}
To enable \textit{SpatialEmb} to support arbitrary microphone topologies, the number of channels \(M\) cannot be fixed. We examine several approaches, both existing and proposed, classifying them into two categories based on the feature fusion methods discussed in Section 2.3.2.

\subsubsection{Spectral Feature Squeezing}
\vspace{-0.2cm}
The arbitrary multi-channel spectral feature is mapped to a single-channel output and then concatenated with the spatial feature, resulting in a fixed output shape of \([2, T, F]\).
\begin{enumerate}
    \item \textbf{Fixed Single-Channel}: Select the spectral feature of the reference channel (the first channel in this work).
    \item \textbf{Random Channel}: Randomly select one channel of the spectral feature.
    \item \textbf{Channel Average}: Average the spectral feature over all channels for each frame.
    \item \textbf{Cross-channel Attention (CCA)}: Apply an attention mechanism across channels for each frame.
\end{enumerate}

\begin{table*}[t]
\centering
\caption{Comparison of various \textit{SpatialEmb} structure with a \textbf{fixed} 8-channel microphone array from different aspects, focusing on efficiency and performance metrics. Efficiency metrics, reported for a 60-second batch during the forward pass in training mode, include their own latency, relative latency to conformer encoder blocks, total FLOPs  (together with conformer encoder blocks), and their own peak memory usage, as well as their relative usage compared to the conformer encoder blocks. Performance metrics are reported based on the Character Error Rate  (CER) for both AliMeeting Eval and Test sets. Version markings are as follows: \textit{S} for small,  \textit{L} for large, \textit{D} for deep, and \textit{Sub} indicates an additional subsampling on the frequency dimension.}
\label{tab:fixed}
\resizebox{\linewidth}{!}{\begin{tabular}{c|c|c|c|c|c|c|c|c|c|c|c|c}
\toprule
\hline

\multirow{2}{*}{Aspect} &\multirow{2}{*}{Module} &\multirow{2}{*}{Version} &\multirow{2}{*}{Spectral Feature} &\multirow{2}{*}{Structure} &\multirow{2}{*}{\shortstack{Output \\Dimensions}} &\multirow{2}{*}{\shortstack{Latency \\  (ms)}} &\multirow{2}{*}{\shortstack{Relative \\ Latency}} &\multirow{2}{*}{\shortstack{FLOPs \\  (G)}} &\multirow{2}{*}{\shortstack{Memory \\  (GB)}} &\multirow{2}{*}{\shortstack{Relative \\ Memory}} &\multicolumn{2}{c}{CER  (\%)} \\ 
& & & & & & & & & & &Eval &Test \\
\midrule
\multirow{7}{*}{C} &\multirow{2}{*}{Conv2d} & \textit{S} &\multirow{12}{*}{201-dim LPS} &\multirow{2}{*}{conv, sub, sub} &16,32,128 &22 &0.21 &262 &0.9 &0.5 &19.76 &23.11 \\
& &\textit{L} & & &64,128,184 &34 &0.32 &349 &1.7 &1.0 &19.91 &23.21 \\ 
\cmidrule{2-3} \cmidrule{5-13}
&\multirow{2}{*}{ConvNext} &\textit{S} & &\multirow{2}{*}{\shortstack{\{conv, next\}, \{sub,\\next\}, \{sub, next\}}}&16,32,128 &48 &0.46 &298 &3.7 &2.2 &19.75 &23.27 \\
& &\textit{L} & & &64,128,184 &100 &0.95 &574 &11.9 &7.0 &18.63 &21.8 \\
\cmidrule{2-3} \cmidrule{5-13}
&\multirow{2}{*}{GRU-Conv2d} &\textit{S} & &\multirow{2}{*}{gru, sub, sub} &16,32,128 &24 &0.23 &262 &2.0 &1.2 &20.67 &24.22 \\
& &\textit{L} & & &64,128,184 &41 &0.39 &372 &6.2 &3.6 &19 &22.16 \\
\cmidrule{1-3} \cmidrule{5-13}
\multirow{5}{*}{\textit{L}} &\multirow{2}{*}{ConvNext} &\textit{D} & &\multirow{2}{*}{\shortstack{\{conv, next\},\{sub, \\next $\times$ 3\}, \{sub, next $\times$ 3\}}} &16,32,128 &60 &0.57 &353 &5.7 &3.4 &18.89 &22.26 \\
& &\textit{L}, \textit{D} & & &64,128,184 &110 &1.05 &838 &18.9 &11.1 &18.02 &21.17 \\
\cmidrule{2-3} \cmidrule{5-13}
&\multirow{2}{*}{GRU-Conv2d} &\textit{D} & &\multirow{2}{*}{\{linear, gru $\times 2$\}, sub, sub} &16,64,128 &23	&0.22 &303	&3.7&2.2 &21.2 &24.73 \\
& &\textit{L}, \textit{D} & & &64,128,184 &54	&0.51	&464	&7.1	&4.2 &18.14	&21.38 \\
\midrule
\multirow{3}{*}{F} &\multirow{3}{*}{Conv2d} &\textit{S} &40-dim Fbank &\multirow{2}{*}{conv, sub, sub} &16,32,128 &\textbf{9} &\textbf{0.09} &\textbf{245} &\textbf{0.5} &\textbf{0.3} &20.48 &24.01 \\
& & \textit{S}&80-dim Fbank & &16,32,128 &10 &0.10 &251 &0.6 &0.4 &19.5 &22.95 \\
& &sub &201-dim LPS &sub, sub, sub &16,32,128 &11 &0.10 &251 &0.7 &0.4 &19.83 &23.3 \\
\midrule
Best &GRU-Conv2d &\textit{L}, \textit{D} &80-dim Fbank &\{linear, gru $\times$ 2\}, sub, sub &64,128,184 &26 &0.25 &330 &3.1 &1.8 &\textbf{17.5} &\textbf{20.91} \\
\hline
\bottomrule
\end{tabular}}
\vspace{-0.5cm}
\end{table*}

\subsubsection{Spatial Feature Expansion}
\begin{figure}
    \centering
    \includegraphics[width=80mm]{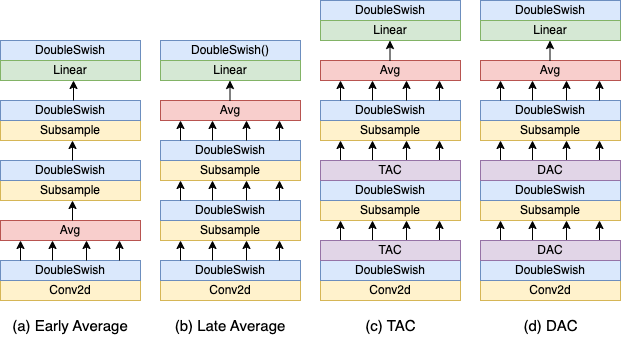}
    \caption{Methods under the "Spatial Feature Expansion" category for arbitrary microphone topology, on top of the \textit{Conv2d SpatialEmb} structure.}
    \label{fig:dac}
    \vspace{-0.5cm}
\end{figure}

\vspace{-0.2cm}
The spatial feature is expanded to the same shape as the multi-channel spectral feature \([M, T, F]\), resulting in \(M\) pairs of spectral and spatial features. They are concatenated, resulting in an output shape of \([M, 2, T, F]\). The \textit{SpatialEmb} module processes each channel independently, maintaining an intermediate size of \([M, C, T, F]\), and averages over \(M\) to obtain a fixed-size representation of \([C, T, F]\). We apply four approaches in this category to the baseline \textit{Conv2d} structure (Fig. \ref{fig:dac}).

\begin{enumerate}
    \item \textbf{Early Average}: The average operation is done right after the first block.
    \item \textbf{Late Average}: The average operation is done right before the \textit{Linear} layer.
    \item \textbf{Transform-Average-Concatenate (TAC)} \cite{yoshioka2022vararray}: For an intermediate representation \(I_m \in \mathbb{R}^{C \times F \times T}\) of the \(m\)-th channel, the output \(O_m \in \mathbb{R}^{C \times F \times T}\) of a TAC module is:
    \begin{equation}
        O_m = [\text{ReLU}(A \cdot I_m); \frac{1}{M}\sum_m \text{ReLU}(B \cdot I_m)]
    \end{equation}
    where \(A\) and \(B\) are linear layers mapping \(C\) to \(C/2\). The concatenated output \(O_m\) remains the same shape as \(I_m\). Finally, an \textit{Average} layer averages \(O_m\) to produce a fixed-dimension output.
    \item \textbf{Proposed: Divide-Average-Concatenate (DAC)}: To reduce computational cost, we simplify TAC to a parameter-free version. We divide \(I_m\) into two halves \([\hat{I}_m; \bar{I}_m]\) with \(\hat{I}_m, \bar{I}_m \in \mathbb{R}^{C/2 \times T \times F}\). The first half \(\hat{I}_m\) retains unique channel information, while the second half \(\bar{I}_m\), representing shared information, is averaged over all \(M\) channels. The output \(O_m \in \mathbb{R}^{C \times T \times F}\) is:
    \begin{equation}
        O_m = [\hat{I}_m; \frac{1}{M}\sum_m \bar{I}_m]
    \end{equation}
    This approach saves computational cost with minimal loss of model capacity.
\end{enumerate}

\vspace{-0.2cm}
\section{Experiments}
\subsection{Data}
\vspace{-0.2cm}
To evaluate our multi-channel multi-speaker ASR system, we utilize the AliMeeting corpus \cite{yu2022m2met}, a challenging dataset featuring Mandarin meeting conversations with multiple speakers. The corpus comprises 104.75 hours of training data, 4 hours of evaluation data, and 10 hours of test data. Each dataset includes several meeting sessions, with each session lasting between 15 and 30 minutes and involving 2 to 4 participants. Unlike the original challenge that permits the use of additional training data, our training is solely based on the far-field 8-channel audio recordings (\textit{Train-Ali-far}), without access to near-field data or microphone topology information, to better simulate real-world conditions.

\subsection{Fixed microphone topology}
\vspace{-0.1cm}
We first determine the best setting for \textit{SpatialEmb} in fixed microphone topology cases without compromising generalizability. To balance the trade-off between performance (CER) and efficiency (computational cost), we vary \textit{SpatialEmb} in three aspects: number of intermediate output dimensions \(C\), number of layers \(L\), and input feature dimension \(F\).

As shown in Table \ref{tab:fixed}, scaling up \textit{SpatialEmb} by increasing channels or layers results in consistent gains for \textit{ConvNext} and \textit{GRU-Conv2d}, but not for \textit{Conv2d}. The intermediate representation within \textit{SpatialEmb} is kept as a 3D tensor \([C, T, F]\), leading to higher computational cost compared to conformer blocks. Therefore, \textit{GRU-Conv2d} is more desirable than \textit{ConvNext} due to lower resource requirements. Additionally, using 80 filter bins (Equation 2) at the entry yields better performance and lower computational cost than the baseline 201-dimensional LPS or a learnable subsampling layer. With these findings, we select \(\textit{GRU-Conv2d}^{L,D}\) as our best \textit{SpatialEmb} structure, improving CER on the \textit{Eval} and \textit{Test} sets from 19.76\%/23.11\% to 17.5\%/20.91\% with an acceptable increase in computational cost.

\vspace{-0.1cm}
\subsection{Arbitrary microphone topology with all channels}

\begin{table*}[t]
\centering
\caption{Comparison of various methods for \textbf{arbitrary} microphone array topology training, along with a fixed 8-channel topology baseline. Please note that we still use all 8 channels for training, although these methods support arbitrary input channels. Metrics are reported the same as in Table \ref{tab:fixed}. Version markings are as follows: \textit{S} for small, \textit{L'} for large with reduced dimension from 64 to 32 on GRU layers, and \textit{D} for deep.}
\label{tab:arbitary}
\resizebox{\linewidth}{!}
{\begin{tabular}{c|c|c|c|c|c|c|c|c|c|c|c}
\toprule
\hline

\multirow{2}{*}{Method} &\multirow{2}{*}{Module} &\multirow{2}{*}{Version} &\multirow{2}{*}{\shortstack{Input \\ Feature} } &\multirow{2}{*}{Structure}&\multirow{2}{*}{\shortstack{Latency \\  (ms)}} &\multirow{2}{*}{\shortstack{Relative \\ Latency}} &\multirow{2}{*}{\shortstack{FLOPs \\  (G)}} &\multirow{2}{*}{\shortstack{Memory \\  (GB)}} &\multirow{2}{*}{\shortstack{Relative \\ Memory}} &\multicolumn{2}{c}{CER  (\%)} \\ 
& & & & & & & & & &Eval &Test \\
\midrule
Fixed 8-chs &\multirow{9}{*}{Conv2d} &\multirow{9}{*}{S} & $[LPS \times 8; SF]$ &\multirow{5}{*}{conv, sub, sub} &22 &0.2 &262 &0.9 &\textbf{0.5} &19.76 &23.11 \\
\cmidrule{1-1} \cmidrule{4-4} \cmidrule{6-12}
Fixed 1-ch & & &\multirow{4}{*}{$[LPS; SF]$} &&22 &0.2 &260 &0.9 &\textbf{0.5} &23.72 &27.02 \\
Random 1-ch& & & & &22 &0.2 &260 &0.9 &\textbf{0.5} &22.06 &25.5 \\
Avg 1-ch& & & & &22 &0.2 &260 &0.9 &\textbf{0.5} &21.28 &24.77 \\
CCA & & & & &66 &0.6 &508 &5.5 &2.1 &20.6 &23.91 \\
\cmidrule{1-1} \cmidrule{4-12}
Early Avg & & &\multirow{4}{*}{$M \times [LPS; SF]$} &conv, avg, sub, sub &38 &0.4 &265 &4.0 &1.5 &20.78 &24.24 \\
Late Avg & & & &conv, sub, sub, avg &47 &0.4 &326 &5.0 &1.9 &20.84 &24.42 \\
TAC & & & &conv, tac, sub, tac, sub, avg &76 &0.7	&337	&7.4	&2.8	&20.04	&23.69\\
DAC & & & &conv, dac, sub, dac, sub, avg &50 &0.5 &326 &6.0 &2.3 &19.6 &22.91 \\
\midrule
DAC &GRU-Conv2d &\textit{L'}, \textit{D} &{$M \times [Fbank; SF]$} &\shortstack{linear, gru $\times$ 2, dac, \\sub, dac, sub, avg} &69	&0.7 &553 &12.6 &4.8 &\textbf{18.18} & \textbf{21.64}\\
\hline
\bottomrule
\end{tabular}}
\vspace{-0.5cm}
\end{table*}

\begin{table*}[t]
\centering
\caption{Results for various arbitrary microphone topology training with random channels selected from 8-channels \textit{Train-Ali-Far} for training and different numbers of channels for inference on Eval and Test sets (\%).}
\begin{threeparttable}

\begin{tabular}{lccccccccc}
\toprule
\hline
\multicolumn{1}{c}{\multirow{2}{*}{Model}}  & \multicolumn{4}{c}{Eval}         &  & \multicolumn{4}{c}{Test}         \\ \cline{2-5} \cline{7-10} 
\multicolumn{1}{c}{}                        & 2-ch & 4-ch & 6-ch & 8-ch &  & 2-ch & 4-ch & 6-ch & 8-ch \\ \hline
Channel Avg + Conv2d     &26.42	&23.31	&22.18	&21.66                        & &29.2	&26.57	&25.52	&25.11\\
TAC + Conv2d  &24.14	&20.22	&19.18	&18.52 & &27.28	&23.49	&22.47	&21.94\\
MFCCA \cite{yu2023mfcca} &25.40	&20.00	&19.50	&19.40 & &26.90	&22.00	&21.50	&21.30 \\
DAC + $\text{GRU-Conv2d}^{L'}$ (best proposed) &\textbf{22.32}	&\textbf{18.46}	&\textbf{17.34}	&\textbf{17.04} & &\textbf{25.7}	&\textbf{21.75}	&\textbf{20.77}	&\textbf{20.32} \\
\hline
\bottomrule
\end{tabular}
\end{threeparttable}
\vspace{-0.5cm}

\label{tab:arb_random}
\end{table*}

\vspace{-0.2cm}
As shown in Table 2, we investigate the best approach for arbitrary microphone topology on a standard \textit{Conv2d} embedding structure, also including a fixed multi-channel result for comparison. Using all 8 channels for training and testing facilitates easy comparison. Among the spectral feature squeezing methods, cross-channel attention (CCA) shows the best performance but at higher computational cost, still leaving a gap to the fixed microphone baseline. Conversely, spatial feature expansion methods show more promising results. Our proposed DAC method not only saves computational cost compared to TAC but also achieves better CER. Combining DAC with our best \textit{SpatialEmb} structure results in 18.18\%/21.64\% CER for an arbitrary microphone topology. The gap to the best fixed microphone topology may be due to reduced GRU layer dimensions to manage high memory cost.

\vspace{-0.2cm}
\subsection{Arbitrary mic topology with random channels}
\vspace{-0.1cm}
We further test \textit{SpatialEmb} on arbitrary microphone topology by varying the number of channels used during training (with permutation) and performing inference on a different number of channels. As shown in Table \ref{tab:arb_random}, our best proposed system using DAC and \textit{GRU-Conv2d} achieves the best performance under all conditions. It significantly outperforms the simple channel average and \textit{Conv2d} embedding baseline and is also better than the best MFCCA system in \cite{yu2023mfcca}. Notably, for the 8-channel inference setting, the model trained with random channels performs even better than the model trained with all 8 channels (17.04\%/20.32\% vs 18.18\%/21.64\% vs 17.5\%/20.91\%). We attribute this gain to the extra randomness introduced during training, which helps prevent the model from overfitting on the relatively small dataset.

\vspace{-0.3cm}
\subsection{Comparison to other systems}
\begin{table}[!htb]
\vspace{-0.3cm}

\caption{Comparison to different systems on Eval and Test sets.}
\centering

\begin{threeparttable}[h]

\begin{tabular}{lccc}
\toprule
\hline
System & Data (hrs) & Eval & Test \\ \hline
$1^{st}$ranking \cite{ye2022royalflush}                     & 10,000       & 19.10 & \textbf{20.10} \\
$2^{nd}$ranking \cite{shen2022volcspeech}                     & 917        & 19.20 & 20.80 \\
\hline
ADL-BF\tnote{$\dagger$}  &105 &20.14	&23.58 \\
GSS\tnote{$\dagger$}  &105 & 18.52 &21.94 \\
MFCCA &105 &19.40  & 21.30 \\
\hline
SpatialEmb &105 & \textbf{17.04} & 20.32 \\
\hline
\bottomrule
\end{tabular}
\begin{tablenotes}
\small
\item $\dagger$: This system is re-implemented by ourselves with the same conformer structure and training data as our system for fair comparison.
\end{tablenotes}
\end{threeparttable}
\label{tab:comp}
 \vspace{-0.5cm}

\end{table}

\vspace{-0.1cm}
Finally, we compare our best system with the state-of-the-art systems mentioned in Section 3, as shown in Table \ref{tab:comp}. Our best proposed \textit{SpatialEmb} system demonstrates superior performance to all these competitors and even surpasses the top-ranking system from the original M2MeT challenge, which used much more training data.

\section{Conclusion}
\vspace{-0.1cm}
In this work, we propose a 1-stage multi-channel, multi-speaker ASR system using \textit{SpatialEmb}. This system encodes multi-channel spectral features and solo spatial features into an embedding for the Conformer encoder to perform ASR on the target speech. We also introduce a parameter-free \textit{divide-average-concatenate} (DAC) approach to support arbitrary microphone topologies, outperforming the fixed microphone version. Our best system, using \textit{GRU-Conv2d} for \textit{SpatialEmb}, sets a new state-of-the-art on the AliMeeting Eval and Test sets with just 105 hours of \textit{Train-Ali-Far}. Future work will focus on large-scale training to develop a universal multi-channel ASR system.

\bibliographystyle{IEEEbib}
\bibliography{refs}

\end{document}